\documentstyle[12pt]{article}
\newcommand {\bi}{\bibitem}
\begin{document}
\thispagestyle{empty}
\vspace*{2.5cm}
\begin{center}
{\Large\bf{SLOPE OF THE ISGUR-WISE FUNCTION }}
\end{center}
\begin{center}
{\Large\bf{ FROM A QSSR CONSTRAINT }}
\end{center}
\begin{center}
{\Large\bf { ON THE $\Upsilon B\bar{B}$ COUPLINGS}}
\end{center}
\vskip 1.0cm
\centerline{{\Large\bf {I.Caprini}}}
\vskip 0.5cm
\centerline {Institute for Atomic Physics,
 POB MG 6}
\centerline {Bucharest, Romania}
\vskip 2.5cm
\nopagebreak\begin{abstract}
\noindent
We derive optimal upper and lower bounds on the slope of the Isgur-Wise
function at the zero recoil point
in terms of the sum of the $\Upsilon B\bar{B}$ couplings,
estimated recently from a $QCD$ Spectral Sum Rule ($QSSR$). The problem is
solved by means of a duality theorem in functional optimization. Optimal
correlations between the slope and the convexity parameters of the
Isgur-Wise function with the same input are also obtained.
\end{abstract}
\newpage\setcounter{page}1

\section{Introduction}
In the last time, there has been some interest in the investigation of the
$b$-number form factor $F(q^2)$ of the $B$-meson
$$<B(p')\vert V^{\mu} \vert B(p)>=(p+p')^{\mu } F(q^2)    \eqno (1.1)$$
where $V^{\mu}=\bar {b}\gamma ^{\mu}b$ and $q=p'-p$. In the
large quark mass limit, this form factor coincides with
the renormalized Isgur-Wise function [1]
$\xi (\omega), ~\omega=v \cdot v'=1-q^{2}/2M_{B}^{2}$,
describing the semileptonic decays of $B$ into $D$ and $D^*$ mesons.
The values of the form factor in the physical region relevant for
these decays  ($ \omega \geq 1$, {\it i.e.}  $q^2<0$ )
 are of interest for the determination of the mixing matrix element
$\vert V_{cb}\vert$ from experimental data. Various models for the Isgur-Wise
function have been proposed in the last time in the frame of the
heavy-quark effective theory ($HQET$). On the other hand, the possibility
of
obtaining model independent bounds on the values of $F(q^2)$, particularly
on its slope at the origin (the charge radius) was outlined for the first
time in [2]. These bounds follow from analyticity, unitarity
and the asymptotic perturbative $QCD$ expansion
for the amplitude of the vacuum polarization due to the $V_\mu$ current.
As noticed in the subsequent papers [3-6], the derivation [2] suffered from
the drawback that it did not include the three upsilon poles present in the
form factor $F$ below the $B\bar B$ threshold.
The problem was correctly treated in [7,8], where bounds
on $F(q^2)$ were derived using as input only the position of the three
upsilon poles below the $B\bar B$ threshold.  Although the methods
are different, the results obtained in [7] and
[8] are both optimal. With a weak form of the unitarity [7],
and using as input only the poles location, known from the experimental
masses of the $\Upsilon$ resonances, the following
bounds on the slope of the Isgur-Wise function were obtained
$$-6.\leq F^{'}(1)\leq 4.1, \eqno (1.2)$$
where the derivative is with respect to $\omega$. The upper bound
is actually not interesting, since it is much larger than the Bjorken upper
bound -0.25 [2].
In [7] bounds on $F'(1)$
were also calculated by assuming that the residua of
$\Upsilon$ the poles are known. Using as input several particular
values for the $\Upsilon B\bar B$ couplings, the conservative lower bound
$$F'(1)\geq-1.5 \eqno (1.3)$$
was proposed in [7], but the authors stress upon the fact that
the only rigorous lower bound on the slope  of the $b$-number form factor
at the origin is (1.2).

It is of interest to know whether this bound can be improved using only
model independent information about the $\Upsilon B \bar B$ couplings.
Recently,  using a $QCD$ Spectral Sum Rule ($QSSR$), Narison [9]
obtained an estimate for the sum of the residua
of the three upsilon resonances $\Upsilon(1S),\Upsilon(2S),\Upsilon(3S)$
situated below the threshold $4M_B^2$. However, in [9]
this estimate was
used only as an indication in choosing several particular values for the
$\Upsilon B\bar B$ couplings. Upper and lower bounds on
$F'(1)$  were then computed with the method presented in [7], for
these specific values of the couplings. No attempt
was made up to now to evaluate the real constraining power of the relation
derived in [9],{\it i.e.} to calculate bounds on the slope of the Isgur-Wise
function in terms of the sum of the upsilon pole residua.

In the  present paper we address this specific problem. We
derive upper and lower bounds on the slope $F'(1)$ using as
input the positions of the
$\Upsilon $ poles and the sum of their residua. We also derive
an optimal inequality relating the slope and the convexity parameters
of the Isgur-Wise function, using the same input.
These are the best model independent theoretical  results which can
obtained without choosing particular values for the $\Upsilon B \bar B$
couplings.

The paper is organized as follows:
in the next Section we derive upper and lower
bounds on the slope $F'(1)$, using
analyticity, unitarity, the $QCD$ one-loop expansion for the polarization
function and the sum of the $\Upsilon B\bar B$ couplings derived from $QSSR$.
We solve the problem by applying
a duality theorem in functional optimization [10,11].
We also generalize the method to include higher derivatives
of the form factor at the origin. The results, expressed
as explicit inequalities involving the quantities of interest, are
investigated numerically in Section 3.
\section{Derivation of the bound}
We consider the $b$-number form factor $F(q^2)$ defined in (1.1), which
is a real-analytic function in the complex plane $t=q^2$, cut along the real
axis from $4M^2_B$ to infinity, except for $N=3$ poles below the $B\bar B$
threshold, corresponding to the states $\Upsilon (1S), \Upsilon (2S)$ and
$\Upsilon (3S)$. Their contribution to the form factor can be parametrized
as [7]
$$F(t)=F(0)+t \sum_{i=1}^{3}{3g_{\Upsilon_{i}B\bar B}f_{\Upsilon_{i}} \over
 M^2_{\Upsilon_{i}}-t-i\epsilon }+......      \eqno (2.1)$$
where $f_{\Upsilon_{i}}$ denote the coupling constants which govern the
electronic widths of the $\Upsilon_{i}$-resonances
$$\Gamma(\Upsilon_{i}\rightarrow (\gamma )\rightarrow e^{+}e^{-})=
f^{2}_{\Upsilon_{i}}M_{\Upsilon_{i}}{4\pi \over 3}\alpha^{2} \eqno(2.2)$$
and $g_{\Upsilon_{i} B\bar B}$ are the coupling constants of the
 $\Upsilon_{i}$
resonances to the $B\bar B$ system. As in [7] we denote by $\eta_{i}$
the product of the above coupling constants
$$\eta_{i}=3g_{\Upsilon_{i}B\bar B}f_{\Upsilon_{i}},~~~ i=1,2,3. \eqno(2.3)$$
By using a $QSSR$ for the vertex function, Narison [9]
estimated recently the sum of these couplings:
$$\sum_{i=1}^{3}\eta_{i}=\alpha =-0.224.\eqno(2.4)$$
We shall take this model independent relation as input in deriving
bounds on the  derivative of the Isgur-Wise function at the origin.
We use also the normalization condition
$$F(0)=1 \eqno(2.5)$$
and the integral inequality [7]
$${n_f\over 12} {1\over 16M^2_B\chi (Q^2)}~ \int^\infty_1~
{(y-1)^{3/2} \over y^{3/2}(y+Q^2/4M^2_B)^2}~\vert
F(4M^2_B y)\vert ^2dy~\leq~1 \eqno(2.6)$$
where $y={t\over 4M^2_B}$. Here
$$\chi (Q^2)~=~{N_c\over 2\pi ^2}\int^1_0{x^2(1-x)^2dx\over m_b^2+
Q^2x(1-x)} \eqno(2.7)$$
 is the $QCD$ one-loop level expression of the
derivative of the polarization function, for spacelike values $Q^2 \geq 0$,
and $N_c$ is the number of colours.
In deriving (2.6) it was
assumed [7] that the unitarity sum for the polarization
function is saturated with the
lowest $B\bar B$ states and that the contribution of the intermediate
states $B^+ B^-, B^0 \bar {B}^0$ and $B_s^0 \bar{B}^0_s$ is the same in
the limit where the light quark mass differences are small.
The stronger version of unitarity used in
[2,6,8], which requires additional dynamical assumptions, beyond $HQET$, will
not be adopted below.

In the inequality (2.6) the spacelike  momentum $Q^2$
enters as a parameter. The dependence
of the results on the choice of $Q^2$, as well as on the $QCD$ sum rule
(dispersion relation) written for the polarization function
 was discussed in [8].
In what follows we shall take the particular value $Q^2=0$.
Then (2.6) becomes
$${5n_f\over 16N_c}~\int^\infty_1~
(y-1)^{3/2} y^{-7/2}~\vert
F(y)\vert ^2dy~\leq~1. \eqno(2.8)$$
We write this inequality in a canonical form by performing the conformal
mapping
$$z~=~{\sqrt {1-y}-1\over \sqrt {1-y}+1} \eqno(2.9)$$
which applies the cut y-plane onto the interior of the unit disk
$\vert z \vert < 1$, such that $z(0)=0$ and the beginning of the
cut, $y=1$ is transformed into $z=-1$. In the new variable the positions
of the $\Upsilon $ poles are denoted by $z_i$
and the residua
of the form factor $F(z)$ at these poles can be written as
$$r_i=(z-z_i)F(z)\vert_{z=z_i}=\eta_iz_i{1-z_i\over 1+z_i}\eqno(2.10)$$
with $\eta_i$ defined in (2.3).
Also, the inequality (2.8) becomes
$${1\over 2\pi}\int^{2\pi}_0~w(\theta)\vert F(\theta)\vert ^2
d\theta~ \leq 1~, \eqno  (2.11)$$
where
$$w(\theta)={5\pi n_f\over 48}cos^4{\theta\over 2}sin{\theta\over 2}\eqno(
2.12)$$
is a nonnegative weight. Of interest for us are the slope and the convexity
parameters of the form factor
$$\rho^2=-F'(1)~,~c={1\over 2}F''(1)~.\eqno(2.13)$$
the derivatives being with respect to $\omega$,
which are related to the derivatives of $F$ with respect to $z$ through
$$F'_z(0)=-8\rho^2~,~F''_z(0)=128c-32\rho^2~.\eqno (2.14)$$

We shall obtain upper and lower bounds on
the derivatives of $F$, using as input the analyticity
of the form factor in the cut $t$-plane, except for the $\Upsilon$ poles,
 and the conditions (2.4), (2.5) and (2.6)
( the last one being written alternatively as (2.11)).

In order to bring the problem to a standard
form, we consider first the function [2,7]
$$\phi(z)~=\phi (0)(1+z)^2 \sqrt {1-z} ,~~~ \phi(0) =
{1 \over 16}\sqrt {{5\pi n_f \over 2N_c}},\eqno(2.15)$$
which is analytic and nonzero inside the unit disk $\vert z \vert<1$
and whose modulus on the boundary is related to the weight $w(\theta )$
given in (2.12) through
$$\vert \phi(\theta)\vert~=~{\sqrt {w(\theta)}}~.\eqno (2.16)$$
Let us define now the new function
$$g(z)=F(z)\phi(z)B(z) \eqno(2.17)$$
where $F$ is the form factor and
$$B(z)~=~\prod ^3_{i=1}{z-z_i\over 1-z_i^* z}. \eqno(2.18)$$
is a product of so-called Blaschke factors [10] (in our case, to a good
approximation, $z_i^*=z_i$). It is clear that the function $g$ is analytic
in $\vert z \vert <1$, the poles of $F$ being canceled by the corresponding
zeros of $B$. Moreover, the constraints (2.4), (2.5) and (2.11)
on the form factor can be written immediately in terms of the function $g$.
Thus, using (2.10) and (2.15) we have
$$g(z_i)=r_i \phi (z_i) \left[ { B(z)\over z-z_i} \right]_{z=z_i}
,\eqno(2.19)$$
and the condition (2.4) becomes
$$\sum_{i=1}^3 \gamma_i g(z_i)=\alpha.\eqno(2.20)$$
Here the weights $\gamma_i$ are defined as
$$\gamma_i=-{1 \over \phi(0)z_i(1+z_i)(1-z_i)^{3/2}\hat {B}_i},\eqno
(2.21)$$
with
$$\hat {B}_i=\left [ {B(z) \over z-z_i} \right ]_{z=z_i}.\eqno(2.22)$$
As concerns the conditions (2.5) and (2.11), they become
$$g(0)=\phi(0) B(0)\eqno (2.23)$$
and, respectively
$$\Vert g \Vert_{L^2}~\equiv~\left\{{1\over 2\pi}\int^{2\pi}_0\vert
 g(\theta)\vert ^2d\theta~\right\}^{1/2}~ \leq~1,\eqno (2.24)$$
where we took into account (2.16) and the fact that the
Blaschke factors have modulus equal to 1 along the unit circle.
Using a standard terminology [10], (2.24) means that the function $g$ belongs
to the unit sphere of the Hilbert space $H^2$.

We express also the derivatives of $g(z)$ at the origin in terms
of the derivatives of $F(z)$, which are further related to the slope and
convexity parameters through (2.14). Using the relations
$$\phi'(0)={3\over 2} \phi (0),~~\phi''(0)=-{\phi(0)\over 4},\eqno(2.25)$$
which follow from (2.15), we obtain
$$g'(0)=\phi(0)B(0)\left[F'(0)+{3\over 2}+{B'(0)\over B(0)}\right]$$
$$g''(0)=\phi(0)B(0)\left\{F''(0)+F'(0)\left[3+2{B'(0)\over B(0)}\right]
-{1\over 4}+3{B'(0)\over B(0)}+{B''(0)\over B(0)}\right\},\eqno(2.26)$$
where we took into account (2.5) and the derivatives of $B$ follow
from (2.18).

We pass now to the calculation of bounds on the derivatives of the
analytic function $g$
in the origin, using as input the constraints (2.20), (2.23) and (2.24).
We first
consider only the slope of the form factor, {\it i.e.} the derivative
$g'(0)$. As will be clear, the generalization to higher
derivatives is straightforward.

In order to solve the problem we shall apply a method based on a norm
minimization over a convex set in $H^2$ and a duality theorem
in functional optimization [10,11]. A similar technique
was applied in [12] for other problems in particle physics.
More precisely, let us consider the
following set of functions:
$$K~=~\left\{ g\in H^2\vert~g(0), ~g'(0)~ {\rm given},~~
\sum_{i=1}^3 \gamma_i g(z_i)=\alpha. \right\}\eqno(2.27)$$
It is clear that $K$ is a convex set in the Hilbert space $H^2$.
In defining it, we took into account the constraint (2.20) and the fact
that $g(0)$ is known, according to (2.23).
In addition, we ascribed to the derivative $g'(0)$, which is the
quantity of interest, a definite value, for the moment arbitrary.

We have still to impose the condition (2.24) and
this will restrict the allowed range of
the unknown parameter $g'((0)$. More precisely, if this parameter is
correctly chosen, all the functions in the set $K$ will have $L^2$ norms
less than or equal to 1. If, on the other hand, $g'(0)$ is outside the
allowed range, the $L^2$ norms of all the functions in $K$ will be greater
than 1, and this will be true also for the smallest of these norms.
It follows that the inequality
$$\min_{g \in K }\Vert g \Vert_{L^2} \leq~1,\eqno(2.28)$$
where the l.h.s. denotes the minimal $L^2$ norm over the set $K$,
describes the optimal range of the parameter $g'(0)$.
As we shall see, the minimum norm (2.28) will depend explicitely on the
specific information
used in the definition of $K$, which consists from the values $g(0),g'(0)$
 and the real number $\alpha$. This will allow us to obtain explicitely
 rigorous upper
and lower bounds on the unknown value $g'(0)$ in terms of the value $g(0)$
and of the parameter $\alpha$.

 Having reduced the
problem to the calculation of a minimum norm, we resort for solving it
to a duality theorem in functional analysis
[10,11]. Usually, a duality theorem relates a
minimization over an abstract space to a maximization in the dual space,
which is often simpler than the original one.  More precisely, if $K$ is a
convex set in the Hilbert space $H^2$ and $g$ denotes
an arbitrary function belonging to $K$, the theorem of interest for us
has the form [10,11]:
$$\min_{g\in K}\Vert g\Vert_{L^2}=\sup_{
G\in L^2,\Vert G\Vert_{L^2}\leq 1}\left\vert -
\sup_{g\in K}{1 \over 2\pi i}\oint_{\vert z\vert =1}
G(z)g(z)dz\right\vert~.\eqno(2.29)$$
Here the supremum in the right hand side is calculated upon all the
complex functions $G$ defined on the boundary of the unit circle, of $L^2$
norms less than 1.  In writing (2.29) we took into account the fact that the
dual of the space $L^2$ is again $L^2$.

We have first to calculate the supremum over $g\in K$ in the r.h.s. of this
equation. In a standard terminology [10] this supremum is called the
{\em support functional} of the convex $K$. Of course, of interest are only
the finite values of the support functional. This clearly restricts the
form of the the functions $G(z)$ taken into account in the functional
optimization. As we mentioned above, the function $G$ entering
the maximization (2.29) is a complex function of $L^2$ norm less than 1,
being in general the boundary value of a function
having an analytic part inside the unit disk, and possible interior
singularities. By residua theorem, these singularities pick up in the
integral (2.29) the values of $g$ at some interior points. It is clear that
these values must be known, according to the
definition of the set $K$, otherwise the supremum upon $g\in K$ would be
uncontrolably large. By applying this argument to the particular
set defined above we see that the functions $G$ yielding a finite
support functional of $K$ have the general form
$$G(z)={a_1\over z}+{a_2\over z^2}+\sum_{i=1}^3{b_i\over z-z_i}+Q(z)
\eqno(2.30)$$
where the residua $a_1,a_2,b_i$ and the analytic function $Q(z)$ are
arbitrary.
By inserting this expression in (2.29) we obtain
$$\min_{g\in K}\Vert g \Vert_{L^2}=\sup_
{G\in H^2,\Vert G \Vert _{L^2}\leq 1}\sup_{g\in K}
\left\vert a_1g(0)+a_2g'(0)+\sum_{i=1}^3b_ig(z_i)\right\vert, \eqno(2.31)$$
where we applied the residua theorem, which implies in particular that the
analytic function $Q$ does not contribute.

It is clear that the supremum with respect
 to $g$ in the r.h.s. of this
equation is finite only if the residua $b_i$ are proportional to the
weigths $\gamma_i$ entering the constraint in the
definition of $K$. We must have therefore
$$b_i=a_3\gamma_i\eqno(2.32)$$
where $a_3$ is an arbitrary parameter, along with $a_1$ and $a_2$.
Then (2.31) becomes
$$\min_{g\in K}\Vert g \Vert_{L^2}=\sup_{\{a_i\}}[a_1g(0)+a_2g'(0)+a_3\alpha],
\eqno(2.33)$$
where the parameters $a_i$ entering the maximization
in the r.h.s. must satisfy the condition that the $L^2$ norm of $G$ is
less than 1. By performing the Fourier analysis of the function $G$ of the
form (2.30) we obtain after a straightforward calculation
 the quadratic constraint
$$(a_1+a_3\sum_{i=1}^3\gamma_i)^2+(a_2+a_3\sum_{i=1}^3\gamma_iz_i)^2+
+a_3^2\sum_{i,k=1}^3
{\gamma_i\gamma_kz_i^2z_k^2\over 1-z_iz_k}\leq 1,\eqno(2.34)$$
which must be satisfied by the parameters $a_i$.

The linear optimization problem (2.33) with the quadratic constraint (2.34)
can be solved immediately, by means of a linear change
of variable which brings this constraint to a diagonal form. The solution
can be written in a compact form, and using
(2.28) we finally obtain the inequality
$$\left\{g'^2(0)+g^2(0)+{1\over \lambda}\left[\alpha-g(0)\sum_{i=1}^3\gamma_i
-g'(0)\sum_{i=1}^3\gamma_iz_i\right]^2\right\}^{1\over 2}\leq 1,\eqno(2.35)$$
where
$$\lambda=\sum_{i,k=1}^3{\gamma_i\gamma_kz_i^2z_k^2\over 1-z_iz_k}.
 \eqno(2.36)$$
Using (2.23) and (2.26) the inequality (2.35) can be written
in terms of $F(0)$ and $F'(0)$, allowing one to calculate upper and lower
bounds on the slope of the Isgur-Wise function using as input the sum ,
$\alpha$,  of
the $\Upsilon$ pole residua. The numerical results are presented in Section 3.

Before closing this section we mention that the technique
applied above can be generalized in a straightforward way as to include
higher derivatives of the form factor in the origin.  This can be done
by a suitable definition of the convex set $K$ and a proper choice of the
function $G$ used in the duality theorem (2.29). Instead of (2.27) we now
define
$$K~=~\left\{ g\in H^2\vert~g(0), ~g'(0)~, ~g''(0)~ {\rm given},~~
\sum_{i=1}^3 \gamma_i g(z_i)=\alpha. \right\}\eqno(2.37)$$
and the function $G$ of (2.30) is replaced by
$$G(z)={a_1\over z}+{a_2\over z^2}+{a_3\over z^3} +
\sum_{i=1}^3{b_i\over z-z_i}+Q(z). \eqno(2.38)$$
A calculation similar to the one presented above leads to the
 following inequality involving the first
two derivatives of the form factor:
$$\left\{{g''^2(0)\over 4}+g'^2(0)+g^2(0)+{1\over
\xi}\left[\alpha-g(0)\sum_{i=1
}^3\gamma_i
-g'(0)\sum_{i=1}^3\gamma_iz_i-{g''(0)\over 2}\sum_{i=1}^3\gamma_iz_i^2
\right]^2\right\}^{1\over 2}\leq 1,\eqno(2.39)$$
where
$$\xi=\sum_{i,k=1}^3{\gamma_i\gamma_kz_i^3z_k^3\over 1-z_iz_k}.
 \eqno(2.40)$$
The inequality (2.39), written in terms of the slope and convexity
 of the Isgur-Wise function by means of (2.23),(2.26) and (2.14),
yields a rigorous correlation between these parameters and the sum
$\alpha$, of the $\Upsilon$ poles residua.   We shall apply it in
the next Section for testing  specific models of the $HQET$.
\section{ Numerical results and conclusions}
In our analysis we took $M_B=5.3~ GeV$ and varied
$m_b$ in the range $4.7-5.3~GeV$.
We first calculate upper and lower bounds on the derivative $F'(1)$, using
(2.35). The results are presented in Table 1, for several values of the
parameter $\alpha$ in the constraint (2.4).
\vskip 1.cm
\begin{tabular}{p{3.cm}|c|c|c|} \cline{2-4}
& $\alpha$  & $F'(1)_{min}$ & $F'(1)_{max}$ \\ \cline{2-4}
&-2.240  & -3.05  & 4.30  \\ \cline{2-4}
&-0.272  & -3.91  & 3.77  \\   \cline{2-4}
&-0.224  & -4.13  & 3.61 \\    \cline{2-4}
&-0.0224 & -4.23  & 3.55  \\   \cline{2-4}
&0.      & -4.23  & 3.53  \\   \cline{2-4}
&0.224   & -4.36  & 3.45  \\   \cline{2-4}
&0.672   & -4.55  & 3.29   \\   \cline{2-4}
&2.240  & -5.15  & 2.59   \\ \cline{2-4}
\end{tabular}
\vskip 0.5cm
\begin{tabular}{l p{10.cm}}
{\it Table 1} :& Upper and lower bounds on the slope of the Isgur-Wise
function for various values of the sum (2.4) of the $\Upsilon $
pole residua.
\end{tabular}
\vskip 0.5cm
In Table 1 we investigated for completeness several values of $\alpha$,
up to ten times larger than the value proposed in [9] and given in (2.4).
Values with opposite sign were considered as well. Taking $\alpha=-0.224$
we obtain
$$-4.13~\leq~F'(1)~\leq~3.61\eqno(3.1)$$
which slightly improve the values (1.2) and
is the best bound on the slope of the $b$-number form factor,
obtained using only model independent information about the residues.

We consider now the inequality
(2.39) relating the slope and the convexity parameters of the Isgur-Wise
function to the sum of the $\Upsilon B\bar B$ couplings.
According to (2.39), the domain of the allowed values
for these parameters is the interior of an ellipsa in the plane $\rho^2-c$.
It is of interest to establish the compatibility
of some particular models of $HQET$ with this rigorous inequality.
For illustration, we consider the following parametrizations proposed in [9]:
$$F(\omega)\simeq 1+F'(1)(\omega -1),$$
$$F(\omega)\simeq exp\{F'(1)(\omega -1)\},$$
$$F(\omega)\simeq \left\{{1\over 2}(1+\omega )\right\}^{2F'(1)},$$
$$F(\omega)\simeq {2\over 1+\omega}exp\left\{[2F'(1)+1]{\omega -1\over
\omega+1}\right\}. \eqno(3.2)$$
These parametrizations, with
the choice $F'(1)\simeq -1.0$,
were used in [9] for a new determination of the mixing angle $V_{cb}$.
They correspond, respectively, to the
following expressions for the convexity parameters:
$$c=0.$$
$$c={1\over 2}[F'(1)]^2$$
$$c={1\over 4}F'(1)[2F'(1)-1]$$
$$c={1\over 2}\left[F'(1)^2-F'(1)-{1\over 4}\right].\eqno(3.3)$$
In order to test the compatibility of the models (3.2) with
the constraint (2.4), we calculate the derivatives $g'(0)$ and
$g''(0)$ using the relations (2.14)
and (2.26) and we insert them in (2.39). The result of this calculation
is given
in Table 2, which indicates the l.h.s. of (2.39) denoted by $LHS$, for several
v
alues of
$\rho^2$ and the parameter $c$ computed according to (3.3) for each
specific model (3.2).
 The two columns correspond to  $\alpha= -0.224$ as
proposed in [9], and also to a value three times larger, respectively.
Unlike the bounds on the slope given in Table 1,
which had a slow variation with $\alpha$, the results are now quite sensitive
to the sum of the pole residua. According to (2.39), a result less than 1
is expected to occur, if the parametrizations (3.2) are consistent with
the constraint (2.4). However,
as seen from Table 2, for $\rho^2=1$  a value
greater than 1 is
obtained for all the models (3.2), which shows that
these models are not compatible with
the relation (2.4) proved in [9]. This throws
some doubt on the determination of $V_{cb}$ made in [9], which was based on
the models (3.2) with the choice $\rho^2=1$. On the other hand, for $\rho^2=
0.5$, the models (3.2) turn out to be consistent with the relation (2.4),
while they fail to be so if the value of $\alpha$ is increased by a factor
of 3.
\vskip 0.5cm
\begin{tabular}{p{2.cm}|c|c|c|c|} \cline{2-5}
&$\rho^2$ & $c$ & $LHS$& $LHS$ \\
& & & $\alpha=-0.224$ &$\alpha=-0.672$ \\ \cline{2-5}
& 1. &0. & 1.65 & 1.95  \\ \cline{2-5}
& 1. & 0.5 & 1.37 & 1.78 \\ \cline{2-5}
& 1. & 0.75 & 1.39 & 1.83 \\ \cline{2-5}
& 1. & 0.875 & 1.44 & 1.89 \\ \cline{2-5}
& 0.5 & 0. & 0.91 & 1.29 \\ \cline{2-5}
& 0.5 & 0.125 & 0.87 & 1.28 \\ \cline{2-5}
& 0.5 & 0.25 & 0.86 & 1.30 \\ \cline{2-5}
& 0.5 & 0.25 & 0.86 & 1.30  \\ \cline{2-5}
\end{tabular}
\vskip 0.5cm
\begin{tabular}{l p{10.cm}}
{\it Table 2 :} & Left hand side of (2.39), denoted as $LHC$, for
 $\rho^2=1$ and $\rho^2=0.5$,
and $c$ computed from (3.3) for the models (3.2). The two columns correspond to
$\alpha=-
0.224$ and to a value three times larger.
\end{tabular}
\vskip 0.5cm

In conclusion, in the present paper we succeeded to fully exploit the
constraining power of the $QSSR$ relation (2.4) between the $\Upsilon
B\bar B$ couplings, derived in [9]. The problem was solved by means of some
powerful techniques in functional optimization.

The results show that the relation (2.4)
has a rather weak constraining power on the slope
of the Isgur-Wise function. In particular, it slightly improved the lower
bound on $F'(1)$ from the value -6, when no information on the
$\Upsilon B\bar B$ couplings is used, to the value  -4.13, which is still very
generous.
On the other hand, the constraint (2.4) is able to produce nontrivial
correlations between the slope and the convexity parameters of the form
factor. The relation (2.39) derived in the present paper turns out
to be an useful tool for
investigating the compatibility of the specific models of the quark theory
with analyticity, unitarity and $QCD$.

\end{document}